\documentclass[article,twocolumn]{revtex4}

\usepackage{graphicx}
\usepackage{amsmath,amssymb,amsfonts}
\newcommand{\be}{\begin{equation}}
\newcommand{\ee}{\end{equation}}
\newcommand{\ba}{\begin{eqnarray}}
\newcommand{\ea}{\end{eqnarray}}
\newcommand{\ban}{\begin{eqnarray*}}
\newcommand{\ean}{\end{eqnarray*}}

\begin{document}

\title{Hot Spaghetti: Viscous Gravitational Collapse}

\author{Berndt M\"uller}
\affiliation{Department of Physics, Duke University, Durham, NC 27708-0305, USA}
\author{Andreas Sch\"afer}
\affiliation{Institut f\"ur Theoretische Physik, Universit\"at Regensburg, D-93040 Regensburg, Germany}
\begin{abstract}
We explore the fate of matter falling into a macroscopic Schwarzschild black hole for the simplified case of a radially collapsing thin spherical shell for which the back reaction of the geometry can be neglected. We treat the internal dynamics of the in-falling matter in the framework of viscous relativistic hydrodynamics and calculate how the internal temperature of the collapsing matter evolves as it falls toward the Schwarzschild singularity. We find that viscous hydrodynamics fails when either, the dissipative radial pressure exceeds the thermal pressure and the total radial pressure becomes negative, or the time scale of variation of the tidal forces acting on the collapsing matter becomes shorter than the characteristic hydrodynamic response time.
\end{abstract}

\maketitle

\section{Introduction}

The fate of matter collapsing into a black hole \cite{Oppenheimer:1939ue} remains one of the great riddles of theoretical physics. In General Relativity the matter encounters a space-time singularity within finite proper time after passing an event horizon, and no further prediction of its evolution is possible. When quantum physics in the semiclassical approximation is added to this picture, the black hole eventually radiates away the mass of the collapsed matter in the form of Hawking radiation \cite{Hawking:1974sw}, but this effect does not address the problem of what happens to the matter that has fallen into the black hole, creating the information loss ``paradox'' \cite{tHooft:1995qox}.

It is generally agreed that the resolution of both problems, the singularity problem and the information loss problem, requires a full quantum theory of gravity beyond the semiclassical approximation. In one widely studied version of quantum gravity, loop quantum gravity, the interior space-time of the black hole ``bounces'' when the energy density reaches $M_P^4$ where $M_P$ is the Planck mass \cite{Modesto:2005zm,Modesto:2009ve}. In this way the singularity is avoided, the collapse is reversed, and the black hole radiates its mass away while the incipient event horizon dissolves smoothly. A conceptual description of this process, including a semiclassical model of the singularity-free evaporation of the black hole, has been presented in ref.~\cite{Hossenfelder:2009fc}. A simple extension of General Relativity that incorporates a limiting curvature has recently been proposed by Chamseddine and Mukhanov \cite{Chamseddine:2016uef}, and its relationship to loop quantum gravity  was explored in ref.~\cite{Bodendorfer:2017bjt}.

In string theory, much work has been devoted to an understanding of the microscopic origin of the Bekenstein entropy of black holes and how information can be encoded at the event horizon. The dynamics of the collapse and ensuing evaporation process is not well understood.  Competing pictures of the black hole as a ``fuzzball'' of highly excited strings \cite{Skenderis:2008qn} or the event horizon as a ``firewall'' \cite{Almheiri:2012rt} that would immediately melt additional incoming matter have been proposed, but their validity remains controversial and unresolved.

Here we focus on the dynamics of matter falling into a black hole before it reaches the region of Planck scale space-time curvature. The study of the viscous heating of matter collapsing into a black hole presented here remains fully within the framework of well established physics. It makes use of the insight that quantum physics imposes a lower bound on the ratio of the shear viscosity $\eta$ to the entropy density $s$ of the matter. The Kovtun-Son-Starinets (KSS) conjecture \cite{Kovtun:2004de} asserts that the lowest value of this ratio is $4\pi\eta/s \geq 1$ for matter governed by any consistent relativistic quantum theory. The  KSS bound is believed to be a strict bound for quantum field theories that admit an Einsteinian gravity dual, but is known to be modestly violated for holographic boundary theories of non-Einstein gravity models \cite{Brigante:2007nu,Brustein:2008cg}. We do not consider the effects of bulk viscosity on the dynamics of collapsing matter, because the bulk viscosity vanishes in the conformal limit, which is usually attained in elementary quantum field theories at high density or temperature.

While the effect of the shear viscosity is negligible in the early stages of the collapse, it becomes relevant in the approach toward the space-time singularity as the gravitational shear forces acting on the collapsing matter grow rapidly. As we will show, the Navier-Stokes approximation to viscous hydrodynamics fails below a certain distance from the singularity, requiring the use of the second-order, causal theory of relativistic hydrodynamics. We obtain analytical solutions to the dynamical equations for shear stress and entropy density in a simple, radially symmetric collapse model neglecting back reaction of the in-falling matter onto the gravitational field, which allow us to explore the effect of viscous heating all the way down to the region where effects of quantum gravity become relevant.

Our paper is organized as follows. First, we briefly remind the reader of the effect of tidal forces on matter falling into a black hole. We next summarize the equations governing the gravitational collapse of matter in the framework of second-order relativistic fluid dynamics. We then study the collapse of a thin shell into a Schwarzschild black hole in a convenient comoving coordinate system and solve the resulting equation for the shear stress in the limit of a constant ratio $\eta/s = (4\pi)^{-1}$, where $\eta$ denotes the shear viscosity and $s$ is the entropy density of the fluid and conclude with some speculations about the implications of our result. 

\section{Spaghettification}

In the final phase collapsing matter is transformed into a quark-gluon plasma. The transformation can either be driven by compressional or viscous heating or, in the absence of a sufficiently high entropy at the start of the collapse, will ultimately be driven by the action of tidal forces.  The destructive shear effect of the tidal forces on matter falling into a black hole is colloquially known as ``spaghettification'' \cite{Hawking:1988qt}. The tidal force compresses in-falling matter in the transverse directions and stretches it in the radial direction. The radial stretching first leads to the rupture of macroscopic solid objects, but can even pull apart atoms and eventually hadrons. 

The radial tidal force over a distance $\Delta r$ on an object with mass $m$ in the gravitational field of a spherically symmetric mass $M$ is:
\be
\Delta F = \frac{mc^2 R}{r^3} \Delta r ,
\ee
where $R = 2GM/c^2$ is the Schwarzschild radius, and $G$ is Newton's constant. Hadrons get disrupted when the tidal stretching force over the size of a hadron exceeds the QCD string tension $\kappa \approx 1$ GeV/fm. For a very rough, order-of-magnitude estimate we set $m \sim \Lambda/c$, $\Delta r \sim \hbar/\Lambda$, $\kappa \sim \Lambda^2c/\hbar$ where $\Lambda \approx 200$ MeV/$c$ is the non-perturbative QCD scale parameter. This yields the condition
\be
\Delta F \sim \frac{\Lambda c R}{r^3} \frac{\hbar}{\Lambda} > \frac{\Lambda^2 c}{\hbar} ,
\ee
which is satisfied for
\be
r < r_h \sim \frac{\hbar}{\Lambda} \left( \frac{\Lambda R}{\hbar} \right)^{1/3} .
\ee
For a 10 solar mass black hole, which has a Schwarzschild radius $R = 30$ km, we obtain $r_h \sim 3\times 10^6~\hbar/\Lambda \approx 3$ nm. 

Matter can still be described in this domain by standard techniques of thermal quantum field theory in curved space-time, because the hadronic disruption distance is much larger than the radial distance at which the space-time curvature reaches the Planck scale. This condition is satisfied as long as
$R/r^3 \ll r_P^{-2}$,
where $r_P \approx 10^{-35}$ m is the Planck length. Inserting the expression for the Schwarzschild radius $R$ we obtain
\be
r \gg r_c = r_p (M/M_P)^{1/3} ,
\label{eq:r-c}
\ee 
with $r_c \approx 10^{-22}$ m for a 10 solar mass black hole.

Whatever the mechanism driving the transformation is, during the final stages of its collapse the in-falling matter forms a quark-gluon plasma, a fluid with the lowest specific shear viscosity known and the greatest ability to flow smoothly under conditions of large shear. Yet, as for any fluid, a hydrodynamic description of the quark-gluon plasma fails when it is forces to respond to changes on time or distance scales that are shorter than the relevant hydrodynamic scales. In this study we will attempt to answer two questions: One is whether the rapidly growing gravitational shear strain on the collapsing quark-gluon plasma leads to a failure of the hydrodynamical description, and if so when; the other is, how hot the plasma becomes before it either enters a non-hydrodynamic domain or reaches the domain where quantum gravity effects become relevant.

\section{Viscous collapse}

In order to avoid the analytical complications of solving the full nonlinear equations governing gravitational collapse, we consider the collapse of a thin spherical shell of matter into an existing black hole of mass $M$. We follow the article by Adler {\em et al.} \cite{Adler:2005vn} on analytical models of gravitational collapse, which analyzes the collapse of fluid spheres. For spheres or shells of dust (pressureless matter) the particles fall freely in the Schwarzschild geometry; when an equation of state $p = \alpha\varepsilon$ with $\alpha > 0$ is used, the spherical shell needs to have a nonvanishing surface tension for consistency. We will not be concerned here with this aspect but simply consider the fluid dynamics in an interior region of the collapsing shell. Also from now on, we adopt natural units $\hbar = c = 1$ for convenience.

For the effect of viscosity in gravitational collapse we follow Herrera {\em et al.} \cite{Herrera:2008gq}. Adopting, in part, the notation of \cite{Herrera:2008gq} we use the following form of the metric (note the opposite overall sign):
\be
ds^2 = A^2 dt^2 - B^2 dr^2 - (Cr)^2 (d\theta^2 + \sin^2\theta d\phi^2) .
\label{eq:Herrera}
\ee
We also need the shear stress tensor $\pi_{\mu\nu}$, which we write in the form
\be
\pi_{\mu\nu} = \Omega \left(\chi_\mu \chi_\nu - \frac{1}{3} h_{\mu\nu} \right) ,
\ee
where $\chi_\mu$ is a unit vector, $\chi_{\mu}\chi^{\mu}=-1$, along the radial direction and $h_{\mu\nu}$ is the projector on the hypersurface orthogonal to the four-velocity. The nonvanishing components of the shear strain are (see eq.~(9) in ref.~\cite{Herrera:2008gq}):
\be
\sigma_1^1 = \frac{2}{3} \sigma ,
\qquad 
\sigma_2^2 = \sigma_3^3 = - \frac{1}{3} \sigma ,
\ee
where
\be 
\sigma = \frac{1}{A} \frac{d}{dt} \ln \frac{B}{C} 
\label{eq:shear-scalar}
\ee
is the shear scalar. 

The shear stress tensor can be expressed in terms of a single function $\Omega(r,t)$ as:
\be
\pi^1_1 = \frac{2}{3}\Omega, \qquad \pi^2_2 = \pi^3_3 = - \frac{1}{3} \Omega .
\label{eq:pi-ij}
\ee
The relaxation equation of the shear stress (eq.~(50) in ref.~\cite{Herrera:2008gq}) simplifies in the limit considered here. We go beyond the simplest form of the Israel-Stewart theory considered in \cite{Herrera:2008gq} and include the complete set of second-order terms obtained in the so-called 14 moment approximation to kinetic theory \cite{Denicol:2012cn} (see also \cite{Denicol:2014tha}). The equation then reads
\be 
\tau_{\pi} \left( \frac{d\Omega}{dt} +\frac{4}{3} \Omega\Theta + \frac{10}{21} \Omega\sigma \right) 
= - 2 \eta A \sigma - \Omega A ,
\label{eq:IS}
\ee
where $\eta$ is the shear viscosity, $\tau_\pi$ is the thermal relaxation time of the shear stress, and
\be 
\Theta =  \frac{1}{A} \frac{d}{dt} \ln (B C^2) 
\ee 
is the volume expansion scalar. In the comoving coordinates we will be using, $A=1$, and the equations simplify further. 
In the Navier-Stokes (N-S) limit, $\tau_{\pi} \to 0$, eq.~(\ref{eq:IS}) implies
\be
\Omega = - 2\eta\sigma 
\label{eq:N-S}
\ee
and thus
\be 
\pi_{\mu\nu} = - 2 \eta \sigma_{\mu\nu} .
\label{eq:NS-pi}
\ee

The amount of shear ${2\over 3}\sigma^2 = \sigma_{\mu\nu}\sigma^{\mu\nu}$ and volume contraction $\Theta$ in the collapsing fluid depend greatly on the details of the collapse. In the case of the collapse of a self-gravitating homogeneous dust cloud, there is no shear ($\sigma=0$) but rapid contraction governed by a contracting FWG geometry. Joshi {\em et al.} \cite{Joshi:2001xi} give an explicit expression for the case of dust ($p=0$), but the general solution is more complicated. For a thin shell collapsing into a Schwarzschild black hole, there is both strong shear and volume contraction ($\Theta=-\sigma$). Other collapse scenarios probably lie in between these two extremes. 

Here we adopt the following simplified approach: We consider a pre-existing black hole into which a thin shell of matter falls, which perturbs the metric only marginally, so that we can assume the shell to fall in a Schwarzschild background with Schwarzschild radius $R = 2GM$. The Schwarzschild metric in Painlev\'e-Gullstrand form (see eq.~(50) of ref.~\cite{Adler:2005vn}) is given by:
\be 
ds^2 = dt^2 - (\psi dt - dr)^2 - r^2 (d\theta^2 + \sin^2\theta d\phi^2)
\ee
with $\psi = - \sqrt{R/r}$. We assume that the shell falls along the geodesic of a free particle:
\be
r^{3/2} + \frac{3}{2}\sqrt{R}\,t = \frac{3}{2}\sqrt{R}\,t_0 \equiv \rho^{3/2} .
\label{eq:trajectory}
\ee
It is convenient to consider $\rho$ as the comoving radial coordinate. We now transform the metric from the Painlev\'e-Gullstrand form with coordinates $(r,t)$ to comoving coordinates $(\rho,t)$. Using the differential
\be
dr = \sqrt{\frac{\rho}{r}} d\rho +\psi dt ,
\ee
we easily obtain
\be
ds^2 = dt^2 - \frac{\rho}{r} d\rho^2 - \frac{r^2}{\rho^2} \rho^2 (d\theta^2 + \sin^2\theta d\phi^2) .
\label{eq:metric}
\ee
Comparing with (\ref{eq:Herrera}) we read off the following metric functions:
\be
A=1, \qquad B = \sqrt{\rho/r}, \qquad C = r/\rho ;
\ee
with $r(\rho,t)$ given by (\ref{eq:trajectory}). We can now calculate the shear scalar (\ref{eq:shear-scalar}):
\be
\sigma = - \frac{d}{dt} ( \ln r^{3/2} ) = \frac{3\sqrt{R}}{2r^{3/2}} .
\label{eq:sigma}
\ee

It may be tempting to simply adopt the N-S approximation, where the shear stress is given by (\ref{eq:NS-pi}). However, this limit must fail at some point during the collapse, because the shear stress in the fluid cannot instantaneously adjust to the ever more rapidly growing gravitational shear strain. We thus need to solve the relaxation equation for the shear stress (\ref{eq:IS}) with $A=1$.
 
Since the relaxation time $\tau_{\pi}$ is a function of the temperature $T$, the relaxation equation for the shear stress cannot be solved in isolation, but needs to be solved together with the equation governing the growth of the entropy density. The equation for entropy production in a viscous fluid can be found, e.g., in ref.~\cite{Heinz:2005zi}, eq.~(28):
\be
T s^\mu_{;\mu} = T \frac{1}{\sqrt{g}} \partial_\mu (\sqrt{g} s^\mu) = \pi^{\mu\nu} \sigma_{\mu\nu} 
= - \frac{2}{3} \sigma \Omega ,
\label{eq:ds/dt}
\ee
where as usual the semicolon denotes the covariant derivative and 
\be 
g = -\det g_{\mu\nu} = A^2B^2(C\rho)^4\sin^2\theta = \rho r^3 \sin^2\theta
\ee
is the metric determinant. Since $\sigma$ and $\Omega$ have opposite signs this implies entropy growth as dictated by the second law of thermodynamics.  In the comoving frame the spatial entropy current $\vec{s}$ vanishes, and we have $s^0=s$. The left-hand side of the entropy growth equation thus takes on the form
\be
T s^\mu_{;\mu} = \frac{T}{r^{3/2}} \frac{d}{dt} \left( r^{3/2} s \right) .
\ee
For an ultrarelativistic gas in the high temperature limit, we have $s \propto T^3$ and after a few manipulations we find:
\be
T s^\mu_{;\mu} = 3 \frac{s}{\sqrt{r}} \frac{d}{dt} \left( \sqrt{r} T \right) .
\label{eq:s-grow}
\ee

Equations (\ref{eq:IS}) and (\ref{eq:s-grow}) yield two equations governing the evolution of the temperature and the shear stress. It is useful to rewrite them in terms of the entropy density-normalized shear stress $\omega = \Omega/s$, which has dimension of energy:
\ba
\tau_{\pi} \left( \frac{d\omega}{dt} + \omega \frac{d \ln s}{dt} + \frac{4}{3} \omega\Theta +\frac{10}{21} \omega \sigma \right)
= - 2 \frac{\eta}{s} \sigma - \omega
\label{eq:IS2}
\\
\frac{d}{dt} \left( \sqrt{r} T \right) = - \frac{2}{9} \sqrt{r} \sigma \omega .
\label{eq:s-grow2}
\ea
The advantage of this rewriting is that it removes a large part of the temperature dependence from the shear viscosity, in some limits, all of it. Equations (\ref{eq:IS2}, \ref{eq:s-grow2}) can be solved when the time evolution of the shear strain $\sigma$ and the ratio $\eta/s$ are known. For the collapsing shell in the Schwarzschild geometry, $\sigma$ is given by (\ref{eq:sigma}), and we will assume that $\eta/s$ is a constant. 

There are then two different limits of (\ref{eq:IS}) that need to be considered. One is the domain, at large $r$, where the shear strain $\sigma$ is small and slowly varying, and therefore the left-hand side of (\ref{eq:IS}) can be neglected. This is the domain of validity of the Navier-Stokes approximation. The other is the domain where $\sigma$ is large and varying rapidly (though still on time scales large compared to $\tau_\pi$), where the second term on the right-hand side of (\ref{eq:IS}) can be neglected. We shall denote the radius at which the transition between these two domain occurs by $r_\mathrm{m}$. Finally, as we shall see below, there is a third domain at the smallest radii where the rate of change of tidal forces exceeds the inverse hydrodynamic response time and the collapsing matter dissolves into a collection of nearly free-streaming quanta. We shall denote the radius where this transition occurs by $r_\mathrm{g}$.

\section{Navier-Stokes Limit}

We first consider the Navier-Stokes (N-S) limit ($r > r_\mathrm{m}$):
\be
\omega = \frac{\Omega}{s} = - 2 \frac{\eta}{s} \sigma ,
\label{eq:omega-NS}
\ee
and (\ref{eq:s-grow2}) takes the form
\be 
\frac{d}{dt} \left( \sqrt{r} T \right) = \frac{4}{9} \frac{\eta}{s} \sqrt{r} \sigma^2 = \frac{\eta}{s} \frac{R}{r^{5/2}} .
\ee
Using $dt = - dr \sqrt{r/R}$ at fixed comoving position $\rho$, we obtain
\be
\frac{d}{dr} \left( \sqrt{r} T \right) = - \frac{\eta}{s} \frac{\sqrt{R}}{r^2} ,
\ee
which has the solution
\be
T(r) = T_0 \sqrt{\frac{r_0}{r}} + \frac{\eta}{s} \sqrt{\frac{R}{r}} \left( \frac{1}{r} - \frac{1}{r_0} \right) .
\label{eq:T-NS}
\ee

The N-S approximation is applicable when the viscous correction to the stress tensor is small compared with the thermal stress. Using (\ref{eq:pi-ij}) and the ultrarelativistic form of the thermal equilibrium entropy, $s_{\rm th}=aT^3$, the radial pressure assumes the form:
\be
P_{\rm r} = \frac{1}{4} T s_{\rm th} + \frac{2}{3} \omega s
= \frac{1}{4} T s_{\rm th} - 2 \frac{\eta}{s} \sqrt{\frac{R}{r^3}} s .
\ee
Like other thermodynamic variables, the entropy factor $s$ in the second term contains thermal equilibrium and viscous contributions. Since the second term already represents the viscous contribution to the radial pressure, which is assumed to be small in the N-S limit, it is consistent to approximate the entropy factor $s$ in the second term by the leading contribution, the thermal entropy $s_{\rm th}$. This yields
\be
P_{\rm r} = \frac{1}{4} s_{\rm th} \left( T - 8 \frac{\eta}{s} \sqrt{\frac{R}{r^3}} \right) .
\ee
The condition for the validity of the N-S approximation thus becomes
\be
T \gg 8 \frac{\eta}{s} \sqrt{\frac{R}{r^3}} .
\label{eq:NS-cond}
\ee
According to (\ref{eq:T-NS}), and neglecting the viscous contribution, this condition is satisfied as long as
\be
r \gg r_\mathrm{m} \approx 8 \frac{\eta}{s} \sqrt{\frac{R}{r_0}} \frac{1}{T_0} .
\label{eq:r-NS}
\ee
It is easy to confirm that in this region the viscous correction to the temperature is, indeed, small.

\section{Second-order viscous hydrodynamics}

Since we are interested in the behavior approaching the singularity, we now consider the opposite case where the growth rate of the shear strain $\sigma$ is too fast for the shear stress to adjust instantaneously. In this domain ($r < r_\mathrm{m}$) it is possible to neglect the second term on the right-hand side of (\ref{eq:IS2}). Again using $dt = - dr \sqrt{r/R}$ at fixed comoving position $\rho$, introducing the dimensionless shear stress $\bar\omega = \omega/T$, and inserting the Schwarzschild expressions for $\sigma$ and $\Theta$, the shear stress relaxation equation becomes:
\be
\tau_{\pi} \left( \frac{d\bar\omega}{dr} + \bar\omega\frac{d\ln(sT)}{dr} + \frac{9}{7} \frac{\bar\omega}{r} \right) 
= \frac{3}{r} \frac{\eta}{sT}  .
\label{eq:omega-relax}
\ee
In order to make progress, we need to fix the temperature dependence of the relaxation time $\tau_{\pi}$ of the shear stress. Here we adopt the parametrization
\be
\tau_{\pi} = \beta \frac{\eta}{s T} .
\ee
The proportionality constant has the value $\beta = 5$ in the conformal Boltzmann gas limit \cite{Denicol:2012cn}, and $\beta = 2(2-\ln 2)$ in the strongly coupled super-Yang-Mills theory \cite{Baier:2007ix,Natsuume:2007ty}. Specializing to the conformal limit $s \propto T^3$, the equations for $\bar\omega$ and $T$ become: 
\ba
\frac{d\bar\omega}{dr} - \frac{5}{7} \frac{\bar\omega}{r} + \frac{4}{3} \frac{\bar\omega^2}{r} &=& \frac{3}{\beta r} ,
\label{eq:domega-dr}
\\
\frac{d \ln T}{dr} + \frac{1}{2r} &=& \frac{\bar\omega}{3r} ,
\label{eq:dT-dr}
\ea
where we used the second equation to eliminate $d(\ln T)/dr$ from the first equation. Remarkably, the two equations no longer depend on the value of $\eta/s$ nor on the value $R$ of the Schwarzschild radius. 

Equation (\ref{eq:domega-dr}) is solved by a constant $\bar\omega$ that satisfies the condition 
\be
\frac{4}{3} \bar\omega^2 - \frac{5}{7} \bar\omega - \frac{3}{\beta} = 0 .
\ee 
The negative solution of this equation for $\beta = 5$ is
\be
\bar\omega = \frac{15}{56} - \sqrt{\left(\frac{15}{56}\right)^2+\frac{9}{20}} \approx -0.454 ;
\ee
for the strongly coupled SUSY theory we find
\be
\bar\omega = \frac{15}{56} - \sqrt{\left(\frac{15}{56}\right)^2+\frac{9}{8(2-\ln 2)}} \approx -0.698 .
\ee

We can now use (\ref{eq:dT-dr}) to solve for $T(r)$:
\be
T(r) = T_1 \left( \frac{r_1}{r} \right)^b 
\label{eq:T-r}
\ee
with
\be
b = \frac{1}{2} - \frac{\bar\omega}{3} .
\ee
The expression for $\omega(r) = \bar\omega\, T(r)$ grows more slowly for $r \to 0$ than $\sigma \sim r^{-3/2}$, confirming our assumption that the second term on the right-hand side of (\ref{eq:IS2}) can be neglected in this domain. 

The solution (\ref{eq:T-r}) of second-order viscous hydrodynamics assumes that the time scale of change of the gravitational shear strain $\sigma$ is longer than the hydrodynamic response time $\tau_\pi$. The rate of change of $\sigma$ is given by
\be
\tau_g = \frac{\sigma}{d\sigma/dt} = - \frac{2}{3} \frac{r}{dr/dt} = \frac{2}{3} R \left( \frac{r}{R} \right)^{3/2} .
\label{eq:tau-g}
\ee
The constraint $\tau_\pi < \tau_g$ is satisfied when $r > r_\mathrm{g}$ with
\be
r_\mathrm{g}^{\frac{3}{2}-b} = \frac{3\beta}{2} \frac{\eta}{s} \frac{\sqrt{R}}{T_1 r_1^b} .
\label{eq:r-g}
\ee

\section{Solution Matching}

As discussed above, the N-S approximation applies for large values of $r$ where the gravitational shear strain is small. The limiting solution of the second-order theory that we derived in the previous section applies for small values of $r$ where the shear strain is large and growing rapidly. In order to match the two solutions we choose the radius $r_{\rm m}$ where the pressure anisotropy $\xi$ coincides for the two solutions. Using (\ref{eq:pi-ij}), $\xi$ is in both cases given by the expression
\be 
\xi(r) \equiv \frac{T_1^1}{T_2^2}
= \frac{\frac{1}{4}sT + \frac{2}{3}\Omega}{\frac{1}{4}sT - \frac{1}{3}\Omega}
= \frac{1+\frac{8\omega(r)}{3T(r)}}{1-\frac{4\omega(r)}{3T(r)}} .
\ee
Using the expression (\ref{eq:omega-NS}) together with (\ref{eq:sigma}) and the ideal fluid limit of the temperature (\ref{eq:T-NS}) we obtain for the N-S solution:
\be
\xi_{\rm NS}(r) \approx \frac{1-8\frac{\eta}{s}\sqrt{\frac{R}{r_0}}\frac{1}{T_0r}}{1+4\frac{\eta}{s}\sqrt{\frac{R}{r_0}}\frac{1}{T_0r}} . 
\ee
For the causal viscous hydrodynamics solution with $\omega(r) = \bar\omega\, T(r)$ the anisotropy is
\be
\xi_{\rm IS}(r) = \frac{1+\frac{8}{3}\bar\omega}{1-\frac{4}{3}\bar\omega} ,
\ee
which is time independent. The two expressions coincide when 
\be
r = r_{\rm m} \approx -\frac{3}{\bar\omega} \frac{\eta}{s}\sqrt{\frac{R}{r_0}}\frac{1}{T_0} .
\label{eq:r-m}
\ee
It is reassuring that the coefficient $-3/\bar\omega$ in (\ref{eq:r-m}) is numerically of similar size as the coefficient in (\ref{eq:r-NS}). 

The parameters of the matched second-order solution are then related to those of the N-S solution as
\be
T_1 \approx T_0 \sqrt{\frac{r_0}{r_\mathrm{m}}} .
\label{eq:T1}
\ee
We still need to check whether the assumption underlying viscous hydrodynamics, that the viscous corrections to the stress tensor do not exceed the pressure, are satisfied. Using the results just obtained, we find for the shear stress scalar:
\be 
\pi^1_1 = \frac{2}{3} \Omega = \frac{2}{3} \bar\omega s T ,
\ee
while the thermal equilibrium pressure is given by $P_{\rm eq} = \frac{1}{4} s T$. For our two limiting cases of the relaxation time coefficient, we have:
\ba
\frac{2}{3} \bar\omega &\approx & -0.303 \qquad [\beta \approx 5] ,
\\
\frac{2}{3} \bar\omega &=& -0.465 \qquad [\beta = 2(2-\ln2)] ,
\ea
which means that the positivity of the radial pressure is violated in both cases. This result shows that the validity of the formalism of second-order viscous hydrodynamics is, at best, marginal, and that higher-order corrections should be taken into account. However, it is worth noting that the second-order theory represents a vast improvement over the N-S approximation, which fails completely upon approach to the singularity as we saw in the previous section.

\begin{figure}[htb]   
\includegraphics[width=0.95\linewidth]{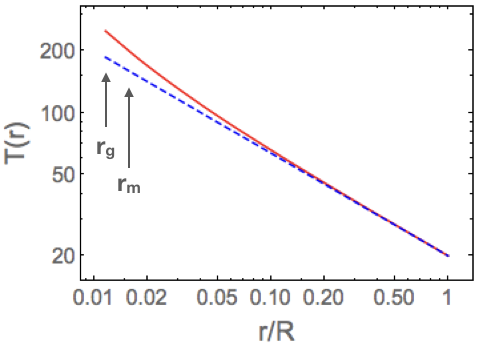}
\caption{The temperature function $T(r)$ (in units of $R^{-1}$) for the selected parameters. The arrows shows the matching radius $r_{\rm m}$ and the radius $r_\mathrm{g}$ where viscous hydrodynamics fails. The solid (red) line shows the solution for viscous hydrodynamics; the dashed (blue) line shows the solution for ideal hydrodynamics.}
\label{fig1}
\end{figure}

In Figure \ref{fig1} we show the complete solution for $T(r)$ for the parameters $r_0/R = 1$, $\beta = 2(2-\ln 2)$, and $T_0 = 20/R$.  The figure shows the temperature in viscous hydrodynamics (solid red line) in comparison with that found in ideal hydrodynamics (dashed blue line). The arrows indicates the matching radius $r_{\rm m}$ and the radius $r_\mathrm{g}$ where the hydrodynamic description fails. Figure \ref{fig1} shows that the effect of viscosity on the temperature in the domain of validity of the N-S approximation is small.

Quite generally, one finds that $r_\mathrm{g} \approx r_\mathrm{m}$ implying that second-order hydrodynamics only slightly extends the extent of validity of hydrodynamics beyond the domain of validity of the Navier-Stokes approximation. This can be seen as follows: Setting $r_1 = r_\mathrm{m}$ in (\ref{eq:r-g}) we can write the equation as
\be
\left( \frac{r_\mathrm{g}}{r_\mathrm{m}} \right)^{\frac{3}{2} - b} = \frac{3\beta}{2} \frac{\eta}{s} \frac{\sqrt{R}}{T_1 r_\mathrm{m}^{3/2}} ,
\ee
where $T_1 = T(r_\mathrm{m})$. On the other hand, combining (\ref{eq:r-NS}) and (\ref{eq:T1}) yields
\be
T_1 r_\mathrm{m}^{3/2} \approx 8 \frac{\eta}{s} \sqrt{R} .
\ee
Inserting this result into the previous equation, we obtain
\be 
\left( \frac{r_\mathrm{g}}{r_\mathrm{m}} \right)^{\frac{3}{2} - b} \approx \frac{3\beta}{16} ,
\ee
which is slightly less than unity for the two values of $\beta$ considered here ($\beta = 5$ and $\beta = 2(2 - \ln 2) \approx 2.61$).

It is instructive to consider some semi-realistic examples. We choose the following parameters: $R = 30$ km, $r_0/R = 1$, $\beta = 2(2 - \ln 2)$, $\eta/s = 1/4\pi$, and consider the following two initial temperatures: $T_0 \approx 0.25~\mathrm{meV}$ (temperature of the cosmic background radiation) and $T_0 = 100~\mathrm{MeV}$ (10 percent of the proton rest mass), assuming substantial pre-heating of the matter during its accretion into the vicinity of the event horizon. The matching radius $r_\mathrm{m}$ and temperature $T_\mathrm{m} = T(r_\mathrm{m})$ are then found as shown in Table \ref{Tab1}.
\begin{table}[htb]
\centering
\begin{tabular}{|c|c|c|}
\hline
$T_0$ & $r_\mathrm{m}$ & $T_\mathrm{m}$ \\
\hline
0.25 meV & 0.5 mm & 1.1 eV \\
100 MeV & 1.3 fm & $2.8\times 10^8$ GeV \\
\hline
\end{tabular}
\caption{Radial distance $r_\mathrm{m}$ and temperature $T_\mathrm{m}$ at the distance where the Navier-Stokes approximation fails, because the radial pressure of the collapsing matter becomes negative. The hydrodynamic model fails shortly afterwards.}
\label{Tab1}
\end{table}
Both the radial distances and the temperature fall well into the domain where quantum gravity effects can be ignored.

The question remains what happens at distances $r < r_\mathrm{g} \lesssim r_\mathrm{m}$ where the hydrodynamical description fails. One possibility is that the hydrodynamical model needs to be replaced by a kinetic description of interacting quanta in phase space for $r < r_\mathrm{g}$ that can account for the fact that the relevant response times of the medium are shorter than the hydrodynamical response times. Another possibility is that the continuous fluid model fails as the radial pressure becomes negative for $r < r_\mathrm{m}$ and medium mechanically ruptures (cavitates) in the radial direction into increasingly smaller domains. Yet another possibility is that the increasingly anisotropic quark-gluon plasma spontaneously develops growing domains of partially coherent color fields that modify the transport equations \cite{Asakawa:2006tc,Asakawa:2006jn}. Possibly more than one of these scenarios is relevant.

\section{Conclusion}

We have explored the fate of matter falling into a macroscopic Schwarzschild black hole for the simplified case of a radially collapsing thin spherical shell for which the back reaction of the geometry can be neglected. We have treated the internal dynamics of the infalling matter in the framework of viscous relativistic hydrodynamics and calculated how the internal temperature of the collapsing matter evolves as it falls toward the Schwarzschild singularity. We found that viscous hydrodynamics fails when either, the dissipative radial pressure exceeds the thermal pressure and the total radial pressure becomes negative, or the time scale of variation of the tidal forces acting on the collapsing matter becomes shorter than the characteristic hydrodynamic response time. We denoted the first radius by $r_\mathrm{m}$, and the second radius by $r_\mathrm{g}$, and found that $r_\mathrm{g} \lesssim r_\mathrm{m}$.

We found that the temperature $T$ of a shell of matter collapsing into a black hole grows as a power of the radius parameter on its approach toward the quantum gravity domain that shields the classical singularity. Since the temperature of the matter remains far below the Planck scale, it can be described by conventional methods of thermal quantum field theory. We found, however, that the pressure anisotropy becomes large in the small-$r$ region where the Navier-Stokes approximation fails. As even the validity of causal second-order viscous hydrodynamics in this domain is questionable, more work will need to be done before definitive conclusions can be reached. An approach that treats the anisotropy of the stress tensor in all orders, such as anisotropic hydrodynamics \cite{Florkowski:2010cf,Bazow:2013ifa} or a kinetic approach \cite{Florkowski:2017olj}, may be better suited to describe this region. 

Another interesting approach, especially for the treatment of a collapsing, strongly coupled quark-gluon plasma, would be a holographic description of the collapse. In such an approach the space-time metric (\ref{eq:metric}) of the black hole represents the boundary geometry of a five-dimensional space-time with a black brane, and the hydrodynamical evolution is recovered from the infrared near-boundary fluctuations around the black-brane background \cite{Kovtun:2003wp}. Many other future research directions come to mind: One would be the numerical solution of the full set of Einstein equations combined with the relaxation equation for the dissipative stress tensor, including the back reaction of the metric to the collapsing matter. By modifying the equation for the gravitational field in such a way that a singularity is avoided, it would be possible to describe the entropy generation during a gravitational bounce.

\section*{In memoriam}

This article is dedicated to the memory of Walter Greiner, our teacher, advisor and mentor during early parts of our careers. Walter was an inspiring scientist and teacher, who taught all that came into close contact with him that the curiosity of the scientific mind should know no bounds. In this, he led by example. It was Walter's mantra that scientific inquiry should not impose limits on itself by sticking to generally accepted approaches to a given problem or established ways of thinking about it. 

For much of his life he was fascinated by the question how the vacuum -- empty space -- responds to external forces, such as strong fields. In his later years as a scientist, Walter was particularly curious about black holes, {\em i.e.} strong gravitational fields. He and his collaborators explored theories that would avoid the existence of black holes \cite{Hess:2010pba,Greiner:2014bjl}. While we do not follow this idea here, we recognize that the question, what kind of objects black holes exactly are, still remains an area of inquiry. 

In our article we address a very specific question: how matter falling into a Schwarzschild black hole is viscously heated by the tidal forces it encounters. In doing so, we connect with two of Walter's lifelong interests: black holes and relativistic hydrodynamics of matter under extreme conditions. We believe that in doing so the article would, at the very least, have captured his interest.

{\em Acknowledgments:} This work was supported in part by grants from the U.S. Department of Energy (DE-FG02-05ER41367). AS thanks the Nuclear and Particle Physics Directorate of Brookhaven National Laboratory for their hospitality. We thank Ulrich Heinz for helpful comments on the correct form of relativistic second-order viscous hydrodynamics and Bill Zajc for several useful comments on an earlier version of our manuscript. We also thank the organizers of the FIAS International Symposium on Discoveries at the Frontiers of Science for the invitation to present a lecture.

\end{document}